\documentclass{article}
\usepackage{graphicx}
\usepackage{latexsym}
\usepackage{amsmath}
\usepackage{amssymb}
\newcommand\scalemath[2]{\scalebox{#1}{\mbox{\ensuremath{\displaystyle #2}}}}
\newcommand{\be}{\begin{equation}}
\newcommand{\ee}{\end{equation}}
\newcommand{\bea}{\begin{eqnarray}}
\newcommand{\eea}{\end{eqnarray}}
\title{ Classical Einstein-Langevin equation and proposed applications }
\author{Seema Satin  \\
\small Dept. of Physics, Tamkang University, Taipei, Taiwan (ROC)\\
\small Dept. of Physical Sciences , Indian Institute for Science Education and 
Research , Mohali, India.\\
\small satin@iisermohali.ac.in}
\vspace{10pt}
\date{ }
\begin{document}
\maketitle
\begin{abstract} 
We propose a Classical Einstein Langevin equation, for
 certain applications in Astrophysics and Cosmology. 
 The domain of applications and the formulation are quite different
 than its semiclassical counterpart ,which is an active area of research and 
 inspires one to carry out this new formalism.
  The field of study can be seen to  emerge out of well
 established ideas and results in  Stochastic Processes in Newtonian Physics
and the Physics in curved spacetime. This is an effort to combine ideas from 
the two areas to get meaningful and new physical results in astrophysics and 
cosmology.
A brief calculation, to demonstrate the contribution
of stochasticity and induced fluctuations to the spacetime metric,  for static
spherically symmetric relativistic stars by heuristic solution 
of the Classical Einstein Langeving equation is given . 
The applicability of the proposed formalism can have a wider expanse
than is mentioned in this article. 
\end{abstract}

Pacs:{04.20.-q, 04.20.Cv, 04.40.-Dg,05.10.Gg,05.40.-a, 98.80Jk,98.80.Bp} 

\section{Introduction}
In recent years, the semiclassical theory of
 stochastic gravity has been taking shape  \cite{BeiL} \cite{Bei2} and is
 finding interesting applications as proposed for
 black hole physics  \cite{suk} \cite{seema} and cosmology \cite{cosverd}.
This relies on including the quantum stress tensor fluctuations in the 
semiclassical Einstein equation, and looking upon its induced effects therein
\cite{Ver,Ver2,nic}.

In another attempt, \cite{moffat} for developing a stochastic theory
 for gravity, a different approach  of incorporating the 
stochastic effects has been proposed. This raises
 a question about smearing out singularities in spacetimes of interest. 
 However the basic difference
 there, lies in  the way stochasticity is introduced. The approach that we
 take up here, is that of introducing stochasticity in a more physically
 relevant fashion using the Langevin formulation . This is in terms of
randomness of the stress tensor itself, however the applications for our
formalism are quite specific and have to be addressed clearly in terms of
the physical picture of randomness.

The theory of classical Brownian motion \cite{B1} \cite{FP} and
 elaborately formulated Semiclassical Stochastic Gravity \cite{BeiL}
 \cite{Bei2}  as mentioned above, naturally
 seem to direct one towards raising a query about possible formulation 
 of a meaningful theory of Classical Stochastic Gravity. 

In what we propose, the modified Einstein equation includes the first order
fluctuations of the classical stress tensor for cases where, the stress
 tensor can be treated as a random variable. This is shown in subsequent
 sections, and has different developments and applications, than
the semiclassical Einstein-Langevin equation.  
 Physical insights and direct applicability to relativistic astrophysics and
 cosmology make such a development quite desirable at this stage.

Perturbations of relativistic stars is a vast and well established area,
the initial developments  \cite{ bardeen, chandra2, chandra3,
chandra4, chandra5, chandra6} started with 
perturbing the metric and analysis of different modes of oscillations in
stellar structure and instabilities. There have been elegant and detailed
 developments since, and spherical  and non spherical perturbations have
 been studied in the context of rotating relativistic stars \cite{fried}
 as well .
 Currently due
 to the detection of gravitational waves \cite{kip0} this area has gained 
 more importance in observational as well as theoretical studies. Though radial
perturbations do not give rise to gravitational waves, nor do the 
scalar perturbations of the metric, any basic and concetually new developments
for perturbations of stars, would naturally motivate one to  begin with the
 simplest cases as that of radial perturbations. We in this article treat as 
the first case of our study to be radial perturbations. Once the basic
formalism as that given in this article is establised, we would give a
similar treatment for non-radial perturbations for different models of 
stars in a full general relativistic set up. 
 
In what we intend to study, the perturbations of the metric would be 
induced ones (not intrinsic )and  sourced by the matter which they are
 composed of.  Hence the
stress tensor is of utmost importance to our analysis, and thus our treatment
is different than the above mentioned approaches. 
The endeavour here, is to extend the the well established analysis of the
perturbations, to include stochastic effect and thus leading to the study of
 the statistical 
properties of the perturbed spacetime. As we plan to
eventually  address slowly rotating and rotating stars in due course
of time,  we would eventually look into stability and collapse 
issues and the  behaviour of statistical correlations therein. With this 
plan in mind, in this first article, we deal with a very simple case, which
can be treated as toy model, to show the  correspondences between
source of stochasticity and the metric pertubations. 
\section{Domain of valid applications: The randomness of the classical stress
tensor}
For astrophysical objects which can be described by giving different models
of the stress-energy tensor (namely perfect and imperfect fluids, classical 
fields etc), the statistical averages of  the stress tensor, are of
 interest. 

In a relativistic star modeled by perfect
fluid, the microscopic particles of the fluid collide frequently, such that
their mean free path is short compared to the scale on which the density
changes. A mean stress tensor in this case can be defined. An observer
moving with an average velocity $u^\alpha$ (four velocity), giving the mean
velocity field of the fluid will see the collisions randomly  distribute
the nearby particle velocities, so that the particle distribution appears
locally isotropic \cite{fried}. The stress tensor can then be treated as a 
random variable and the source of stochasticity in such cases, are the
 collisions of microscopic particles. This has been suggested in \cite{Bei1}.
Neutron star  dynamics, stability issues etc, require a 
detailed knowledge of the stars' microphysics. 
 
The source of these fluctuations may necessarily be the microphysics 
encompassing the quantum phenomena in the interior of the stars, which we may
 be able to partially capture in  fluctuations of the classical 
 variables.  
 It is important to mention here that this analysis is very different from 
 that of the  semiclassical case, where the stochasticty is based on specific 
quantum fields coupled to the spacetime and does not have 
implications to classical counterpart, even though the expectation of
the quantum stress tensor is treated classically in such cases. 

 It may also be  possible  to extend this study consistently, for a very
 different scenario in cosmology, where we treat the FRW metric with
a perfect fluid model for matter.
 The relevant scales of interest there are expected to  decide the 
fluctuations of the stress tensor and the randomness.

 For such cases, one can then phenomenologically put in the fluctuations of
 $T^{ab}$ in the Einstein Equation . This specifies  noise, thus giving it a
 form of a Classical Einstein-Langevin equation. 
\section{The Classical Einstein-Langevin Equation}
The simplest form of  E-L equation can be written as, 
\be \label{el}
G^{ab}[g+h](x) = T^{ab}[g+h](x) + \tau^{ab}[g](x) 
\ee
where  the  $\tau^{ab}[g]$ is the classical stochastic entity, and is 
defined by  $ \tau^{ab}(x)= (T^{ab}(x) - <T^{ab}(x)>)$ of  the unperturbed
 background. It satisfies the condition $\nabla_a \tau^{ab}=0$.  
 The covariant derivative is taken w.r.t the background metric
$g^{ab}$ . This ensures that the Einstein Langevin equation
is covariantly conserved. The term $\tau^{ab}$ gives the equation a stochastic
form and is thus defined only through its expectation value. For the equation to
be meaningful we assume  $< \tau^{ab}(x) > = 0$, which holds for a Langevin
 type noise. 
 The perturbations that are  induced by these fluctuations form the solution
of this equation.

  The magnitude of fluctuations thus
 defined, needs to be small enough to fulfill the criteria for validity of such
 a treatment.
We are thus interested in seeking the effect of these fluctuations on the
statistical properties of the fluid variables and spacetime metric . 
  This can be obtained by solving the above equation formally
for the perturbations $h_{ab}$ of the metric $g_{ab}$.  
 We shall aim for our future work to develop methods for solutions, which
are quite invovled and would depend on specific models.  

 In other applications where the classical
fields associated with the body are of interest, one can take average of the
stress tensor over these  e.g electromagnetic, magnetic or 
electric field, or classical scalar fields in general, associated with the body.

It may be worth mentioning the difference of the averages taken here and
the expectations in semiclassical theory. 
A quantum stress tensor expectation $<\psi|\hat{T}^{\mu \nu}|\psi>$ 
over certain quantum states, as in
 semiclassical Einstein Equation may be treated as classical 
 \cite{holland}.  Also the issues of regularization etc, related
to  the quantum stress tensor make the semiclassical theory very involved in
mathematical developments  regarding formulations and solutions of the
corresponding Einstein Langevin equation. 
The averages of the classical stress tensor that we consider here are
fundamentally different and quite simple.  Here the underlying Physics is in a 
different domain, so this should not be confused with the avegares in the
 semiclassical case. Similarly, the applications of the semiclassical and
 classical stochastic gravity do not overlap. 

Thus equation (\ref{el}) above  can  take the following form, 
\be
G_{ab}[g](x) + \delta G_{ab}[h](x) =  T_{ab}[g](x) + \delta T_{ab}[h](x)
 + \tau_{ab}[g](x)
\ee
which reduces to
\be
 \delta G_{ab}[h](x) =  \delta T_{ab} [h](x) + \tau_{ab}[g](x)
\ee
 We assume the stochastic term to be of the following form, as it
describes the Langevin noise.
\be
< \tau_{ab}(x)> = 0  , < \tau_{ab}(x) \tau_{cd}(x')> = N_{abc'd'}(x,x') 
\ee 
where $ab$ corresponds to $x$ and $c'd'$ to $x'$. The fluctuations
 denoted by $\tau^{ab}$ can be written as
\be
\tau_{ab}(x) = T_{ab}(x) - < T_{ab}(x)>
\ee 
as mentioned earlier.
Thus the two point correlation is given by

\be \label{noise}
<\tau_{ab}(x) \tau_{cd}(x') > = < T_{ab}(x)T_{c'd'}(x') > - <T_{ab}(x)>
 <T_{c'd'}(x')>
\ee
Here  $<T_{ab}>$ is the statistical average of the classical stress 
tensor, since this is treated as a random variable itself. 
The bitensor $N_{abcd}(x,x')$ decribes the two point correlation and has
the following properties.
\begin{enumerate}
\item
\be
 N_{abcd}(x,x') = N_{cdab}(x',x)    
\ee
This is clear from  eqn. (\ref{noise}). 
\item
\be \label{eq:noisecon}
 \nabla_a N^{abcd}(x,x') = \nabla_b N^{abcd}(x,x') = 
\nabla^c N_{abcd}(x,x') = \nabla^d N_{abcd}(x,x')= 0 
\ee
This follows from covariant conservation of $T^{ab}(x)$. 
\end{enumerate}
Such a contribution is ignored in deterministic treatment of perturbations. 
\section{ Induced metric perturbations for relativistic  star}
The formalism for obtaining perturbations of stars around equilibrium
 confinguartion, lies at the core of study of oscillations, stability issues
and critical points of collapse of massive stars.

Any new development in the area, starts with spherically symmetric stars and
 their perturbations. Since we intend to  give a basic framework here, 
as the first exercise, we begin with the simplest case that of a static
configuartion as the background spacetime. 

  One should consider the example worked out here as a toy model.

\subsection{The basic framework}
A static spherically symmetric  spacetime in Schwarzchild coordinates is of
 the form
\be \label{eq:spher}
 ds^2 = - e^{2 \nu(r)} dt^2 + e^{2 \lambda(r) } dr^2 + r^2 d \Omega^2
\ee   
The stress-energy tensor for a perfect fluid is given by 

\be \label{eq:stress1}
T_{ab} = (\epsilon+ p) u_a u_b + g_{ab} p 
\ee
where the four-velocity $u^a = e^{-\nu}(1,0,0,0)$.

We shall discuss the  spherical (radial) perturbations of such a system here
and reserve  the analysis of general perturbations, which includes 
non-spherical cases for later work. 

For the above case of the metric, the perfect fluid stress tensor describing
 the interior has the explicit form 
\be \label{eq:stress2}
T_{ab} =\left(
\scalemath{0.8}{
\begin{array}{cccc}
- e^{2\nu} \epsilon & 0 & 0 & 0  \\
  0  & e^{2 \lambda}p & 0 & 0 \\
0 & 0 &  r^2 p  & 0 \\
0 & 0 & 0 & r^2 \sin^2 \theta p 
\end{array}
}
\right)
\ee
Accordingly, the unperturbed components of field equation are given as
\bea
G_{tt} = 8 \pi T_{tt} & : & \nonumber \\
 - e^{2(\nu- \lambda)} (\frac{1}{r^2} - \frac{2}{r} 
\lambda ' )& + & \frac{e^{2 \nu}}{r^2} =  8 \pi e^{2 \nu}\epsilon \label{eq:1}\\
G_{rr} = 8 \pi T_{rr} & : & (\frac{1}{r^2} + \frac{2}{r} \nu' ) 
- \frac{e^{2 \lambda}}{r^2} = 8 \pi e^{2 \lambda} p \label{eq:2} \\
 e^{ 2 \lambda} G_{\theta \theta} = 8 \pi e^{2 \lambda}  T_{\theta \theta } 
& : & \nonumber \\
& &  \nu '' + \nu'^2 - \nu ' \lambda ' + \frac{1}{r} (\nu ' - \lambda ' )
= 8 \pi e^{2 \lambda} p 
\eea

The perturbed potentials $\delta \nu$, $\delta \lambda$ and fluid variables
 $\delta p$ and $\delta \epsilon $  as that of perturbed pressure and density
  characterize the perturbed system.   
\subsection{The perturbed equations with stochastic contributions}
 \label{sec:pertpot}
Perturbations in a static system, call for time dependence of the
perturbed quantities and fluctuations. For radial perturbations of the
quantities involved here as elaborated below, we assume the time
dependence to be of the form
\be
\delta \lambda(r,t) = \delta \lambda(r) e^{-i\omega t}
\ee  
(and similarly for $\delta \nu$ , $ \delta p$, $\delta \epsilon$ and other
 fluctuating quantities) 
where $ \omega$ is the factor related to the frequency of oscillations. 
This  is inspired by similar kind of time dependence of radial
modes of oscillations in general relativistic stars \cite{kipnotes}. However,
in what follows we will not go into mode analysis of the oscillations and 
related issues, and proceed with the radial
part in order to solve for the Einstein -Langevin equation. 

The perturbed equations, including the fluctuations of the 
random stress tensor corresponding to the above Einstein equations are given
 by ( we will need the first two here) 
\bea
 \delta G_{tt} = 8 \pi \delta T_{tt}(x) + \tau_{tt}(x)  & : &  \nonumber \\
\delta \lambda (\frac{1}{r^2} - \frac{2}{r} \lambda ' )  + \frac{1}{r}
 \delta{\lambda'}  =  4 \pi e^{2 \lambda} (\delta \epsilon  + \tau_{tt})
\label{eq:delam} \\
 \delta G_{rr}(x) = 8 \pi \delta T_{rr}(x) + 8 \pi \tau_{rr}(x) 
 & : &  \nonumber \\
\frac{1}{r} \delta \nu' - \delta \lambda(\frac{1}{r^2} + \frac{2}{r} \nu')
  = 4 \pi e^{2 \lambda} ( \delta p  + \tau_{rr}) & & \label{eq:delnu} \\
\eea

The rest of the perturbed equations can be obtained similarly. In what follows, we deal only with the 'r' dependence of the perturbations for simplicity and
 as shown earlier keep the exponentiated 't' dependence part separate as
it is separable for the static background.

To solve the above equations, we would use along with the above, the 
perturbed Euler equation
\be
\delta [\nabla_a T^{ab} ] = 0 
\ee
From $\delta {\nabla_a T^{at} } =0 $ it follows that $ \delta \epsilon =
 -  \delta \lambda (\epsilon + p) $. 
We further impose a linear
equation of state which describes the perfect fluid constituting
the interior of the massive star, given by
\be
p = w \epsilon
\ee
and perturb this, assuming $w$ to be constant for our case so that
\be \label{eq:p}
\delta p = w \delta \epsilon ; \mbox{ } \delta p = - w(\epsilon + p)
 \delta \lambda
\ee
Putting the above the in the first two perturbed Einstein- Langevin equation,
\begin{eqnarray}
 & & \delta \lambda (\frac{1}{r^2} - \frac{2}{r} \lambda ' )  + \frac{1}{r}
 \delta{\lambda'}  =  4 \pi e^{2 \lambda} (- (\epsilon + p) \delta \lambda
 + \tau_{tt}) \label{eq:lam1} \\
 & & \frac{1}{r} \delta \nu' - \delta \lambda(\frac{1}{r^2} + \frac{2}{r} \nu')
  = 4 \pi e^{2 \lambda} ( - w(\epsilon+ p) \delta \lambda  +
 \tau_{rr}) \label{eq:delnu}
\label{eq:nu1}
\end{eqnarray}
\subsection{Model of Noise}
The model of noise that we use in the above, decides the stochastic behaviour
of the perturbations of the metric. 

Noise is defined by (\ref{noise}), since we consider  a static system  here,
 the stress tensor itself  $T_{ab}(x) $ does not depend  on 't', but 
introducing randomness in the system calls for a time dependence of
the fluctuations.
This in turn introduces, random motion in the fluid particles, such that the
 spherical symmetry of the star is still maintained and the background metric
still treated as static. This special case can be modeled as follows:
\be
 \tau_{ab}(x)  = \tau_{ab}( \vec{x}) e^{-i\omega t} = (T_{ab}(\vec{x}) -
<T_{ab}(\vec{x}) > ) e^{-i\omega t}
\ee
The noise kernel then reads
\be
<\tau^*_{ab}(x) \tau_{cd}(x')>  = <\tau_{ab}(\vec{x}) \tau_{cd}(\vec{x}')>
e^{i\omega(t-t')} 
\ee
where averages are taken over spatial sector only and thus
\be
N_{abcd}(x,x')=  N_{abcd} (\vec{x}, \vec{x}') e^{i \omega(t-t')} =
 <\tau_{ab}(\vec{x}) \tau_{cd}(\vec{x}')> e^{i \omega(t-t')}
\ee
For this special case of noise kernel the conservation is valid only
 for the spatial part, given by
\be
\nabla_a N^{abc'd'}(\vec{x},\vec{x}') = \nabla_b N^{abc'd'} (\vec{x},\vec{x}')
 = \nabla_{c'} N^{abc'd'}(\vec{x},\vec{x}') = \nabla_{d'} N^{abc'd'}(\vec{x}
,\vec{x}') =0
\ee

Thus the  components of noise kernel correpsonding to
the case that we deal here are given by 
\nopagebreak
\bea
& & <\tau_{tt}(r)\tau_{t't'}(r')>  =  e^{2(\nu(r)+ \nu(r'))} \mbox{Cov}[ 
\epsilon(r) \epsilon(r')] \nonumber \\  
& & <\tau_{rr}(r)\tau_{r'r'}(r')>  =   e^{2(\lambda(r)+ \lambda(r'))} 
\mbox{Cov}[p(r)p(r')]  \nonumber \\
& & <\tau_{\theta \theta }(r)\tau_{\theta' \theta'}(r')>  =
  r^2 r'^2 \mbox{Cov}[p(r)p(r')] \nonumber \\ 
& & <\tau_{\phi \phi}(r)\tau_{\phi' \phi'}(r')>  =  r^2 r'^2 \sin^2 
\theta \sin^2 \theta' \mbox{Cov}[p(r)p(r')] \nonumber \\
& & <\tau_{tt}(r)\tau_{r'r'}(r')>  = -e^{2(\nu(r)+ \lambda(r'))}
 \mbox{Cov} [\epsilon(r) p(r')] \nonumber \\   
& & <\tau_{tt}(r)\tau_{\theta' \theta'}(r')> =  -e^{-2(\nu(r))}r'^2
 \mbox{Cov} [\epsilon(r) p(r')]  \nonumber \\
& & <\tau_{tt}(r)\tau_{\phi' \phi'}(r')>  = - e^{2(\nu(r))}r'^2
 \sin^2 \theta' \mbox{Cov}[\epsilon(r) p(r')] \nonumber \\ 
& & <\tau_{rr}(r)\tau_{t't'}(r')> = - e^{2(\nu(r')+ \lambda(r))}
 \mbox{Cov} [p(r)\epsilon(r')] \nonumber \\    
& & <\tau_{rr}(r)\tau_{\theta' \theta'}(r')>  = e^{2(\lambda(r))}
r'^2  \mbox{Cov}[p(r)p(r')]  \nonumber \\ 
& & <\tau_{rr}(r)\tau_{\phi' \phi'}(r')> =  e^{2(\lambda(r))}r'^2
 \sin^2 \theta'  \mbox{Cov}[p(r)p(r')] \nonumber \\ 
 & & <\tau_{\theta \theta}(r)\tau_{\phi' \phi'}(r')> = r^2 r'^2
  \sin^2 \theta'  \mbox{Cov}[p(r) p(r')] \nonumber \\
& & <\tau_{\theta \theta}(r)\tau_{t't'}(r')> = e^{2(\nu(r'))}r^2 
\mbox{Cov}[p(r)\epsilon(r')]  \nonumber \\ 
& & <\tau_{\phi \phi}(r)\tau_{t't'}(r')>  = -e^{2(\nu(r'))}r^2
 \sin^2 \theta \mbox{Cov}[p(r)\epsilon(r')] \nonumber \\
& &  <\tau_{\theta \theta}(r)\tau_{r'r'}(r')> = e^{2(\lambda(r'))}r^2
 \sin^2 \theta \mbox{Cov}[p(r) p(r')] \nonumber \\
& & <\tau_{\phi \phi}(r)\tau_{\theta' \theta'}(r')>  = r^2 r'^2
 \sin^2 \theta  \mbox{Cov}[p(r) p(r')] \nonumber \\ 
& & <\tau_{\phi \phi}(r)\tau_{r'r'}(r')>  = e^{2 \lambda(r')}r^2 
 \sin^2 \theta  \mbox{Cov}[p(r) p(r')] \nonumber \\ 
\eea 
Using this we attempt to obtain solutions of the Einstein Langevin equation
 in terms  of two point correlations of perturbed potentials $\delta \lambda$
 and $\delta \epsilon$.

 It may certainly raise
a query if these can be related to the semiclassical noise kernel as obtained
 in semiclassical stochastic gravity. In order to give an indication
to this, one could first investigate the connection between the
 classical stress tensor here and the quantum stress tensor used in
semiclassical Einstien equations, thus relating pressure and density
inside the model, with quantum states. The pressure here (say for
 a neutron star) arises due to neutron degeneracy, and thus
 has quantum origin. It is open to further investigations how one
 would relate the fluctuations in the two cases, and which quantum
stress tensor should be used to get correct results. Also the
quantum stress tensor for the given case should appropriately
describe the matter of the interior of the star. It would be our endeavour
to work this out in a separate article addressing these issues, investigating
if there can be such a connection between the two cases, in exact
theoretical terms.  
\subsection{ The  two point correlations of  potentials}
Solving equations (\ref{eq:lam1} ) and (\ref{eq:nu1}) 
 to get expressions for $\delta \nu $ and $\delta \lambda$ it follows that
\begin{eqnarray}
\delta \lambda(r) & = & 4 \pi e^{-\int f_1(r) dr} \int e^{\int f_1(\hat{r}) d 
\hat{r}} e^{2 \lambda(r')} r' \tau_{tt}(r') dr'
\end{eqnarray}
where 
\[
f_1(r) = ( \frac{1}{r} - 2 \lambda' + 4 \pi r e^{2 \lambda} (1 + w) \epsilon 
\]
 and 
\begin{eqnarray}
\delta \nu(r) & = & - 4 \pi \int \{ f_2(r'') e^{\int f_1(\bar{r}) d \bar{r}}
\int e^{\int f_1(\hat{r}) d \hat{r}} e^{2 \lambda(r')} r' \tau_{tt}(r') dr' \}
dr'' \nonumber \\
& & + 4 \pi \int e^{2 \lambda(r')} r' \tau_{rr} (r') dr'
\end{eqnarray}
where
\[ f_2(r) = 4 \pi r e^{2 \lambda} r w(1+w) \epsilon + 2 \nu' - \frac{1}{r}) \]
Thus one can clearly see that $<\delta \lambda> =0 $, $ <\delta \nu> =0 $, as
expected.

The two point correlation of the potential perturbations $\delta \nu$ and
 $\delta \lambda$ can be obtained by using the above two equations and
putting in the model of noise from the previous section. 
We present two of these correlations, namely, $<\nu(r) \nu(r')>$ and
 $<\lambda(r) \lambda(r)>$ here. Rest of combinations can be obtained on the
 similar lines.
\begin{eqnarray}
< \delta \lambda(r_1) \delta \lambda(r_2) > & =  & 16 \pi^2
 e^{- (\int f_1(r_1) dr_1 + \int f_1 (r_2) dr_2) } \int \int e^{( \int 
f_1(\hat{r}_1) d \hat{r}_1  + \int f_1 ( \hat{r}_2 ) d\hat{r}_2 )}
\nonumber \\
& &  e^{2 ( \lambda(r_1') + \nu(r_1') + \lambda(r_2') + \nu(r_2')) } Cov[ \epsilon(r_1') \epsilon(r_2') ] dr_1' dr_2'
 \label{eq:covlam}
\end{eqnarray}
\begin{eqnarray} \label{eq:covnu}
& & < \delta \nu(r_1) \delta \nu(r_2) >  =  16 \pi^2 \int \int \{ f_2(r_1'')
 f_2(r_2'') e^{- ( \int f_1( \bar{r}_1) d \bar{r}_1 + \int f_1( \bar{r}_2) 
d \bar{r}_2 )}  \nonumber \\
& & \int \int e^{( \int f_1(\hat{r}_1) d \hat{r}_1 + \int f_1( \hat{r}_2)
d \hat{r}_2 )} e^{2( \lambda(r_1') + \nu(r_1') + \lambda(r_2') + \nu(r_2'))}
r_1' r_2' \nonumber \\
& & Cov[ \epsilon(r_1') \epsilon(r_2') ] dr_1' dr_2' \} dr_1'' dr_2''
+ 16 \pi^2 \int \int e^{4(\lambda(r_1') + \lambda(r_2') )} r_1' r_2' 
 \nonumber \\
& & Cov[ p(r_1') p(r_2')] dr_1' dr_2' + 16 \pi^2 \int \{ f_2 (r_1'') 
e^{- \int f_1(\bar{r}_1) d \bar{r}_1 }
\int  \int e^{\int f_1(\hat{r}_1) d \hat{r}_1} e^{4 \lambda(r_2')} 
 \nonumber \\
& & e^{2( \lambda(r_1')+ \nu(r_1'))} r_1' r_2' Cov[ \epsilon(r_1') p(r_2')]
 dr_1' dr_2' \} dr_1'' + 16 \pi^2 \int \{ f_2(r_2'') 
e^{- \int f_1 (\bar{r}_2) d \bar{r}_2 } \nonumber \\
& & \int \int e^{ \int f_1(\hat{r}_1) d \hat{r}_2 } r_1' r_2' 
 e^{4 \lambda(r_1')} e^{2( \nu(r_2') + \lambda(r_2')) } Cov[ p(r_1')
\epsilon(r_2')] dr_1' dr_2' \} dr_2''
\end{eqnarray}

As we see in the above expressions, the covariances in pressure and
 density in the interior of the relativistic star , decide the stochastic
 contributions to the perturbations of the potentials. Thus the
nature of density and pressure distribution is of utmost importance
 to the Langevin analysis of the system. It can be seen from the above
expressions, vanishing covariances of pressure and density result on vanishing
of the metric pertrubations. Thus the stochastic behaviour of the induced 
metric pertubations here can be seen to arise due to the statistical
 properties of matter density and pressure in the interior of the star. 

 We  would elaborate more on statistical and thermodynamic properties and
a  fluctuation dissipation theorem in an upcoming article. These are deeper
issues, and need to be worked upon separately, while considering  a
solution of the Einstein Langevin equation for more realistic stellar structure.
Thus we see that, the  stochastic contributions from the interior of the star,
 which arise due to microscopic or quantum effects of the highly compact
matter, induce  perturbations in the geometry.  In another
way one can say that, a perturbed geometry 
as a backreaction of these fluctuations, may thus characterise the density 
and pressure distribution profiles inside the star.
The effective (classical) perturbed quantities thus arising due to underlying
 quantum or microscopic phenomena (which may partially be caputured thus)
 without knowing the detailed picture therein at the quantum level and thus
understanding this in a  statistical sense . This gives a mesoscopic
 picture of the interior of a relativistic star, and can be
 explored futher for specific cases by obtaining exact analytical or numerical
solutions of the Einstein Langevin equation. 
\section{Further Directions} 
Here, our purpose has been to establish the formalism theoretically
 and state few basic
consequences of including the effects of fluctuations of the 
classical stress tensor. Further developments are planned in 
upcoming articles, where we would attempt complete solutions of
the classical Einstein Langevin equation and the behavior of the
 perturbations (and their correlations) at critical phases of a collapsing
star using appropriate equation of state to get solutions of the 
Einstein Langevin equation on similar lines. We suspect drastically changing
 behaviour of the
 correlations of perturbations to show up at few of these critical points
of collapse in a dynamical framework modeled by a similar Einstein Langevin
 equation. We would also attempt to analyse the
 growth or decay of these perturbations in the dynamical model of a star and 
related issues.  

 It may be
 interesting to see if, general relativity in this domain(statistical), would
 have at the fundamental level a much richer structure and content, than the 
theory of stochastic processes in Newtonian  Physics.  This can be expected,
 due to
  curved spacetime here, which gets coupled to matter. 
There is indication towards this from other areas, where we see the effect
 of underlying geometry of space as that in fractal structures, which come
 up in a very interesting way \cite{seemaadg}. One can  seek to
analyse the Einstein Langevin equation in terms of Markovian or Non-Markovian
criteria, in addition to the background spacetime stucture affecting the
 same.

Apart from this general interest in stochastic theory development,  the
present article directs one, towards a well defined research program which
 includes the following:
\begin{itemize}
\item A formal solution of the classical Einstein Langevin equation in
terms of general perturbation of the background spacetime metric to be
 obtained by devising methods to do so. Thus solvable models have to be
identified, for which this could be done analytically, along with cases 
 where one may need numerical solutions.  
\item Few cases of interest would be the rotating neutron stars or compact
iastrophysical objects, which are also of interest to Graviational Waves.
\item 
 This analysis  with additional contributions to the the spacetime
 structure and its perturbations, also can be used to study
 instabilities in collapsing bodies.   A formal study towards  
non-nonequilibrium statistical physics can be  provided  in this context.
The instabilities in
gravitational collapse of relativistic stars (viz a neutron 
star or a system of neutron stars), would be amenable to analysis 
 in terms of studying the correlations of the stress tensor
 fluctuations  and their  effects near different critical regions during the
 collapse scenario. 
\item  For the complete gravitational collapse \cite{pankaj1, pankaj2,piran},
 where one necessarily gets
 a singualrity as the end state of the collapse, the effect of these
 stochastic contributions  towards the end states of collapse before the
singularity is reached, can play interesting role in deciding the
stability criteria for the end states. This can modify the present
established results, regarding parameter values that decide the occurrence of 
naked or covered singularity. Also this may enhance the  area by giving it
 a way, to work out issues using statistical physics . 
\item In cosmological spacetime, one may be able to find application of the
classical Einstein Langevin equation on the lines of \cite{star, linde} 
\end{itemize}
The few possible applications of the proposed classical theory of stochastic 
gravity in this article are mentioned, to physically motivate such a
 formulation and bring to notice the directions where this would lead to 
meaningful study. One can always suggest many more sub-areas and
 problems in astrophysics or cosmology, where this may find valid 
 applications.

 Introducing the idea
of such an approach as that of a Langevin equation in classical gravity, is 
just the very first step. This necessarily needs to be followed by the 
appropriate solutions of the Einstein-Langevin equation, including
 mathematical developments for specific cases.
This is our  endeavour in immediate future.

\section*{Acknowledgements}
Seema Satin is thankful to Bei Lok Hu, Sukanta Bose, 
and T. Padmanabhan and Kinjalk Lochan  for helpful discussions.

\end{document}